\documentclass[aps,prb,preprint,showpacs,groupedaddress,amssymb,amsfonts,amsmath]{revtex4} %
\usepackage{amsmath}
\usepackage{amsfonts}
\usepackage{graphicx}
\usepackage{bm}
\usepackage{enumerate}
\usepackage{color}
\begin{document}
\title{Strong terahertz emission from superlattices via Zener
tunneling}

\title{Strong terahertz emission from superlattices via Zener
tunneling}
\author{Peng Han}
\affiliation{Beijing National Laboratory for Condensed Matter
Physics, Institute of Physics, Chinese Academy of Sciences, Beijing
100080, China}
\author{Kui-juan Jin}
\email{kjjin@aphy.iphy.ac.cn} \affiliation{Beijing National
Laboratory for Condensed Matter Physics, Institute of Physics,
Chinese Academy of Sciences, Beijing 100080, China}
\author{Barry C.~Sanders}
\affiliation{Institute for Quantum Information Science, University
of Calgary, Calgary, Alberta T2N 1N4, Canada} \affiliation{Centre
for Quantum Computer Technology, Macquarie University,
        Sydney, New South Wales 2109, Australia}
\author{Yue-liang Zhou}
\affiliation{Beijing National Laboratory for Condensed Matter
Physics, Institute of Physics, Chinese Academy of Sciences, Beijing
100080, China}
\author{Hui-bin Lu}
\affiliation{Beijing National Laboratory for Condensed Matter
Physics, Institute of Physics, Chinese Academy of Sciences, Beijing
100080, China}
\author{Guo-zhen Yang}
\affiliation{Beijing National Laboratory for Condensed Matter
Physics, Institute of Physics, Chinese Academy of Sciences, Beijing
100080, China}

\begin{abstract}
We develop a comprehensive, elegant theory to explain terahertz
(THz) emission from a superlattice over a wide range of applied
electric field,which shows excellent agreement between theory and
experiment for a GaAs/Al$_{0.3}$Ga$_{0.7}$As superlattice.
Specifically we show that increasing electric field increases THz
emission for low fields, then reduces emission for medium fields due
to field-induced wave function localization, and then increases
emission in the high field due to delocalization and Zener tunneling
between minibands. Our theory shows that Zener tunneling resonances
yield high THz emission intensities and points to superlattice
design improvements.
\end{abstract}

\pacs{73.21.Cd, 73.40.-c, 78.47.+p}

\maketitle

The immense value of coherent terahertz (THz) electromagnetic
radiation in many fields, including medical imaging, molecule
recognition, sub-millimeter astronomy, remote sensing, and condensed
matter physics~\cite{Mit99}, has provided the impetus for vigorous
research efforts into creating new sources of reliable and
affordable THz radiation~\cite{Sek99,Asc04} including exploiting
Bloch oscillations in a superlattice (SL)~\cite{Was93}. The SL is
especially promising as a THz source because it is effectively an
artificially engineered semiconductor with tunable parameters and is
thus adaptable. Vigorous experimental
research~\cite{Leo92,Leo03,Shi02,Shi04} has been accompanied by
theoretical advances~\cite{Jin03,Han05,Voj05}, but a proper theory
that explains the physics and the THz emission properties for a SL
THz source over a wide range of electric field has been elusive
until now.

We present a quantitative theory that explains THz emission from a
SL over a wide range of applied electric field, and our theory also
explains qualitatively the rise and fall of emitted THz field
intensity in the low field (LF) regime, the rise of emission in the
medium field (MF) to high field (HF) regime, anomalous resonances
observed in THz emission from a GaAs/Al$_{0.3}$Ga$_{0.7}$As
SL~\cite{Shi04}, and provides insight into the limitations in the
ultrahigh field (UF) regime. Our theory elucidates the subtle
physics of electromagnetic radiation emission in a SL subjected to
high electric fields and, moreover, shows that Zener tunneling
between minibands in the conduction band can be valuable especially
for superior design of THz sources.

Usually theories consider SL behavior in the regime of weak applied electric field
where Zener tunneling (typically deleterious, such as electrical breakdown in semiconductors
due to tunneling to higher bands~\cite{Zen34}), can be ignored~\cite{Jin03}. However
recently we have seen an indication that Zener tunneling between minibands of the
conduction band~\cite{Han05} coincide with the anomalous THz resonances observed in a
GaAs/Al$_{0.3}$Ga$_{0.7}$As SL~\cite{Shi04}. Inspired by this promising connection
between theoretical enhanced Zener tunneling rates and observed THz resonances in
a SL, we have developed a theory for THz emission from a SL subjected
to a range of applied electric field; our theory explains experimental observations and provides
a promising foundation from which to design SLs as versatile THz emitters.

We analyze THz emission intensity~$I$ from a SL in an electric
field~$F$ by subdividing the behavior into four regimes of applied
electric field: (i)~the LF regime corresponds to~$I$ increasing
for increasing applied electric field~$F$; (ii)~the MF regime for
which~$I$ decreases with increasing~$F$; (iii)~the HF regime for
which~$I$ again increases with increasing~$F$ and also
demonstrates resonances; and (iv)~the UF regime (with~$I$
decreasing with increasing~$F$). Previously regimes~(i) and~(ii)
have been separately explained by considering a single miniband
within the conduction band using a Drude model for regime~(i) and
a quantum description for regime~(ii)~\cite{Jin03}; later the Kane
model involving coupled Schr\"{o}dinger equations~\cite{Kan59} was
used to describe the cross-over between regimes~(i)
and~(ii)~\cite{Han06}. More recently the locations of resonances
in regime~(iii) were identified with enhanced Zener tunneling
rates between two minibands within the conduction
band~\cite{Han05} but the characteristics of~$I$ vs~$F$ were not
presented. Here we present a full and complete theory for regimes
(i-iii) that includes all the successes of these prior models and,
furthermore, easily describes all the cross-overs between applied
electric field regimes, shows excellent quantitative agreement
especially at the resonances. Our perturbative two-miniband theory
provides a clear intuitive understanding of THz emission from a SL
in terms of localization of wave functions and properties of both
minibands.

In our model the SL growth direction is along the $x$-axis with $F$ in the positive $x$~direction.
For~$a$ and~$b$ the widths of the well and barrier, respectively, $m$ the total number of
unit cells (one well plus one barrier), $d=a+b$ the unit cell width (and the potential~$V$
periodic over width~$d$),
and~$\hbar$ the reduced Planck's constant, the Hamiltonian for an electron~$-e$
and effective mass~$\mu$ is
\begin{equation}
\label{eq:H}
    H=-\frac{\hbar^2}{2\mu}\frac{\textmd{d}^2}{\textmd{d}x^2}+V(x)+eFx,\;V(x)=V(x+ld).
\end{equation}
The spectrum is subdivided into minibands indexed by~$n$, and we are
interested  in the case of two minibands so $n\in\{1,2\}$. The
eigenfunctions for $F=0$ can be expressed as Bloch functions
$\{\phi_{nk}(x)\}$, with corresponding eigenvalues~$\{E_{nk}\}$ and
Bloch number~$k$ in the Brillouin Zone (BZ). The Bloch functions and
the eigenvalues in the SLs are obtained by using the Kronig-Penney
model \cite{Pan91} with a step size of $d/200$ and the BZ is sampled
by 2000 points. In order to appreciate the role of inter-miniband
tunneling, we study the system with Zener tunneling between
minibands treated by perturbation theory; exact and numerical
methods for solving Hamiltonian $H$~(\ref{eq:H}) tend to obscure the
important role played by tunneling. Furthermore we will see that the
ultrahigh electric field regime, which is characterized by
decreasing~$I$, is where second-order perturbation effects become
important.

The Bloch functions can be used as a basis set for $F \neq 0$ as well.
In this case, a wave function confined to band~$n$ is
$\Phi_{nl}(x)=\sum_{k\in\text{BZ}}a_{nlk}\phi_{nk}(x)$,
with the coefficients $\{a_{nlk}\}$ satisfying a well known relation~\cite{Hou40,Glu04}.
Inter-miniband tunneling is calculated via dual degenerate perturbation theory.
The zeroth-order wave function $\psi^{(0)}_{1l, 2l'}(x)=C_{1l}\Phi_{1l}(x) + C_{2l'}\Phi_{2l'}(x)$
straddles both minibands,
with~$l'$ satisfying the condition that the energy difference
$|\epsilon_{2 l'} - \epsilon_{1 l}|$ is minimized in
$l'\in\{1,\ldots,m\}$, and
\begin{equation}
\label{eq:Cpert}
    \begin{pmatrix}
        H'_{1l,1l}&H'_{1l,2l'}\\H'_{2l',1l}&H'_{2l',2l'}
    \end{pmatrix}
    \begin{pmatrix}
        C_{1l}\\C_{2l'}
    \end{pmatrix}
    =\varepsilon^{(1)}_{1l}
    \begin{pmatrix}
        C_{1l}\\C_{2l'}
    \end{pmatrix}
\end{equation}
with $\varepsilon^{(1)}_{1l}$ the eigenvalue for first-order
perturbation of Eq.~(\ref{eq:H}) and
$H'_{1l,2l'}=\langle\Phi_{1l}(x)|~eFx~|~\Phi_{2l'}(x)\rangle$ the
perturbation matrix element. The first-order wave function is
\begin{align}
    \psi^{(1)}_{1l, 2l'}(x)=\sum_{l^{\prime\prime}\neq
    l}\frac{\langle\psi^{(0)}_{1l^{\prime\prime}, 2
    l'-l+l^{\prime\prime}}(x)|~eFx~|~\psi^{(0)}_{1l,
    2l'}(x)\rangle}{\varepsilon^{(0)}_{1l}-\varepsilon^{(0)}_{1l^{\prime\prime}}}
        \nonumber   \\  \times
    \psi^{(0)}_{1l^{\prime\prime},2 l'-l+l^{\prime\prime}}(x),
\label{eq:1stOrderWavefunction}
\end{align}
with $l^{\prime\prime}\in\{1,\ldots,m\}$ and $\varepsilon^{(0)}_{1l}$ the zeroth order eigenvalue.
In the case of weak tunneling
($C_{1l} \rightarrow 1,\,C_{2l'}\rightarrow 0$), degenerate perturbation
theory merges with non-degenerate perturbation theory.
For first-order perturbation theory, the total wave function is
$\psi_{1l, 2l'}(x)=\psi^{(0)}_{1l,2l'}(x) + \psi^{(1)}_{1l, 2l'}(x)$.
We also perform the calculation to second order in order to separate first-order
effects, which yield desirable resonances, and second-order effects, which we show
limit performance in the ultrahigh-field regime.
Whether the wave function~$\psi_{1l, 2l'}(x)$ is calculated to first- or second-order will
be clear from the context.

\begin{figure}[b]
\centerline{\includegraphics[width=8.5cm]{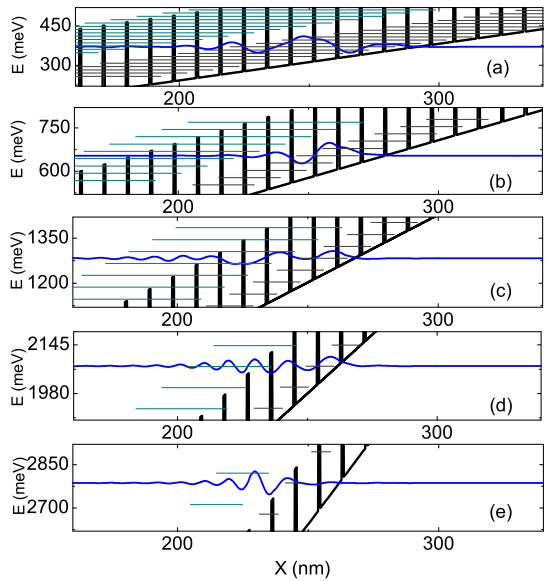}}
\caption{\label{fig1} (Color on-line)
    Plot of Re$\left[\psi_{1l, 2l'}(x)\right]$ (blue line),
    which is the wave function at energy $l,l'$ for applied electric field values
    (a)~$F=14$kV/cm (LF),
    (b)~$F=26$kV/cm (MF),
    (c)~$F=45$kV/cm (HF away from THz resonance),
    (d)~$F=90$kV/cm (HF near THz resonance), and
    (e)~$F=120$kV/cm (UF).
    The horizontal lines are the equally spaced energy levels
    in each of the two minibands (right for the 1st miniband and left for the 2nd miniband).
}
\end{figure}

We plot $\text{Re}\left[\psi_{1l, 2l'}(x)\right]$ in
Fig.~\ref{fig1} for several values of~$F$ to depict characteristics of this wave function
such as localization and overlap with the nearest levels in each of the four $F$ regimes (LF, MF, HF, UF). We see that the wave function can be localized or delocalized and may be confined to one
miniband or overlap two minibands, with these characteristics important for understanding
THz emission.

THz emission intensity depends on the wave function~$\psi_{1l,
2l'}(x)$ and on the applied electric field. The relationship between
THz emission frequency~$\omega_B$ and field strength~$F$ is given by
$\hbar\omega_B=eFd$, which is the energy level splitting in the
miniband. To calculate the total radiative intensity, we adapt the
relation~\cite{Jin03}
\begin{equation}
   F\sum_{l^{\prime\prime}=1}^m |l-l^{\prime\prime}
    |\left|\int_\text{SL}\textrm{d}x\,\psi_{1l, 2
    l'}^*(x)x\psi_{1l^{\prime\prime}, 2
    l'-l+l^{\prime\prime}}(x)\right| ^2
\label{eq:I(F)}
\end{equation}
to include broadening due to scattering; here $l\in\{1,\ldots,m\}$,
$l'$ minimizes $|\varepsilon^{(0)}_{2l'}-\varepsilon^{(0)}_{1l}|$,
and the wave functions implicitly depend on~$F$. For the experiment
to be considered in the present paper, the energy half width of the
pumping laser pulses is approximately 20 meV. Within this energy
interval, the levels in the Wannier-Stark ladders (WSLs) are
populated with equal probability~\cite{Jin03}. Thus the summation
runs over all pairs of WSL eigenfunctions.

It has been demonstrated in Ref.~\cite{Bas91,Tar06} that the
acoustical phonon scattering plays a decisive role in the THz
radiation process. Let $|n,\vec{\rho}\rangle$ be the eigenstate of
the $n$th WSL level, where $\vec{\rho}$ is the two-dimensional (2D)
wave vector. As shown in Ref.~\cite{Bas91}, in the steady state the
electron population on each WSL level is the same, and there will be
no net stimulated emission from the SLs due to the vertical
transition from the initial state $|n + 1,\vec{\rho}\rangle$ to the
final state $|n,\vec{\rho}\rangle$ is perfectly canceled by the
absorption transition from state $|n,\vec{\rho}\rangle$ to state $|n
+ 1,\vec{\rho}\rangle$ that occurs at the same frequency and
intensity without the effect of electron-phonon
scattering\cite{Jin03}. However, in the transition processes involve
the emission or absorption of an acoustical phonon, the in-plane
momentum of the initial state will differ from that of the final
state. Hence, the emitted THz radiation will not be completely
reabsorbed by the system, and the system can have a net gain for the
photon energy $\hbar\omega$ $<$ $\hbar\omega_B$. Because the energy
of an acoustical phonon is much less than the energy of THz photon,
we still treat the $\hbar\omega_B$ as the THz radiation photon
energy in this work\cite{Jin03}.

\begin{figure}[t]
\includegraphics[angle=0,width=7.5 cm]{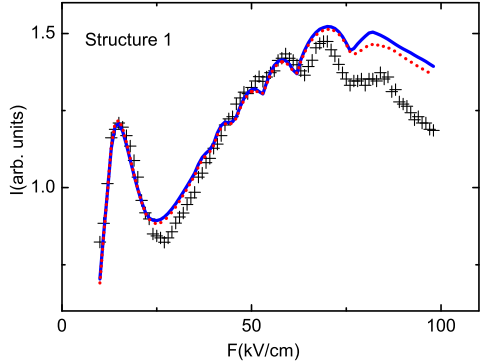}
\caption{\label{fig2} (Color on-line) First-order (solid line) and
second-order (dashed line) perturbation theory results vs
experimental data~[8] (black crosses) for structure 1. }
\end{figure}

\begin{figure}[t]
\includegraphics[angle=0,width=7.5 cm]{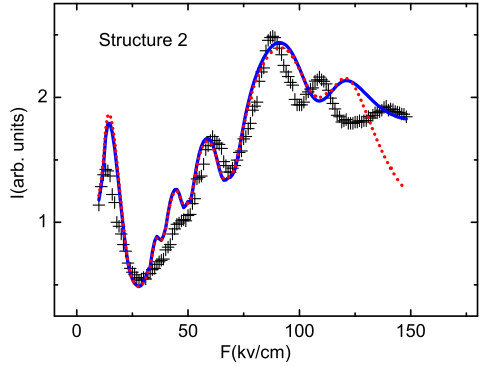}
\caption{\label{fig3}(Color on-line) First-order (solid line) and
second-order (dashed line) perturbation theory results vs
experimental data~[8] (black crosses) for structure 2. }
\end{figure}

Our theory is compared with recent experimental results on two
GaAs/Al$_{0.3}$Ga$_{0.7}$As SL structures~\cite{Shi04}: structure
1, $a=6.4$nm and $b=0.56$nm with $m=73$ (first miniband: 18 to 114
meV; second miniband: 150 to 445 meV) vs structure 2, $a=8.2$nm
and $b=0.8$nm with $m=55$ (first miniband: 19 to 69 meV; second
miniband: 107 to 270 meV). We show our calculated $I(F)$
characteristic for both first- and second-order perturbation
theory in Figs.~\ref{fig2} and ~\ref{fig3} alongside corresponding
experimental results for both structures.

Two fitting parameters are used in our calculations to match
theoretical and experimental curves:
$I_\ell^\text{plot}(F)=\varsigma_0
I_\ell^\text{calc}(F)+\varsigma_1$, with $I_\ell^\text{plot}$ the
plotted~$I$ for structure~$\ell$, $I_\ell^\text{calc}$ the
calculated~$I$, and the fitting parameters $\varsigma_0$ for
rescaling and $\varsigma_1$ for background. 
The choice of both coefficients is jointly determined by least
squares fitting for both sets of experimental data, with excellent
agreement between second-order perturbation theory and experiment.
First-order perturbation results agree for all electric field
regimes except UF, thus revealing the important role of second-order
effects in limiting achievable THz intensity.

Theoretical and experimental radiation intensities $I(F)$ are
depicted in Figs.~\ref{fig2} and ~\ref{fig3} for structures 1 and 2,
with the theoretical curve determined by first- and second-order
perturbation based on Eq.~(\ref{eq:I(F)}).  As
expression~(\ref{eq:I(F)}) ignores scattering, we incorporate the
broadening effects by convolving $I(F)$ with a Lorentzian of
half-width $\Gamma$, which is obtained based on the calculation of
longitudinal optical phonon scattering~\cite{Har05, Tar05, Pet03},
aluminum atom alloy scattering~\cite{Ray92, Pal82} and interface
roughness scattering~\cite{Dha90,Unu06} by using Fermi's golden
rule. Theory and experiment agree well for all field regimes except
UF, and first-order perturbation results also agree for all applied
field regimes except the UF. However note that the theoretical curve
for structure 2 diverges somewhat from the UF intensity peaks of the
experimental curve, which we believe is due to these two peaks
arising from Zener tunneling between the second and third minibands;
furthermore this anomaly is compounded by our least square fitting
parameters $\varsigma_0$, and $\varsigma_1$ truncated for
$F_\text{cut} = 100$ and $F_\text{cut}=120$ kV/cm for structure 1
and 2, respectively. The fitting parameters for the first-order and
second-order perturbation calculation in structure 1 and 2 are given
in table~\ref{t.1}.

\begin{table}
\caption{Fitting parameters for structure 1 and 2} \label{t.1}
\begin{center}
\begin{tabular}{lcr}
\hline \hline
Parameters  & $\varsigma_0$  & $\varsigma_1$\\
\hline
Structure 1 (1st-order)  & 3.667 & 0.316\\
\hline
Structure 1 (2nd-order)  & 3.739 & 0.332\\
\hline
Structure 2 (1st-order)  & 2.598 & 0.164\\
\hline
Structure 2 (2nd-order)  & 2.771 & 0.134\\

\hline
\end{tabular}
\end{center}
\end{table}

The curves clearly exhibit four regimes of behavior according to
the strength of the applied electric field. We see that $I(F)$
rises monotonically for LF, then decreases monotonically  in MF,
and HF is characterized by a trend of $I$ increasing with
rising~$F$ but punctuated by resonances. Finally there is a region
in the UF case with the experimental result for $I$ decreasing
with increasing $F$, which agrees with the second-order
perturbation theory result, but disagreeing with first-order
perturbation result. Our theory explains all these features as
exemplified by Figs.~\ref{fig1}. In Figs.~1(a,b), corresponding to
the low- and medium-field regimes, respectively, the slopes of the
minibands are small, and Zener tunneling between minibands can
thus be ignored. In these two cases the wave functions are
confined to the lowest miniband, but in (a)~the function
progressively spreads out with increasing $F$ whereas it
progressively localizes with increasing~$F$ (due to competition
between localized wave functions) in (b)~\cite{Jin03,Han06}.

The HF regime in Figs.~\ref{fig1}(c-d) exhibits clear effects due to inter-miniband tunneling.
Fig.~1(c) shows steeper minibands than in~\ref{fig1}(a-b), which causes the wave function to overlap
both minibands. Analogous to~(a), the wave function progressively spreads as~$F$ increases,
yielding increasing THz emission with increasing~$F$. However this increasing~$I$ with increasing
$F$ is punctuated with resonances due to Zener tunneling (this connection was speculated
when coincidences between values of~$F$ values for enhanced Zener tunneling and experimentally
observed resonances were demonstrated~\cite{Han05}).

Here we see exactly what is happening at the resonance: the wave
function is localized in both minibands but with significant
overlap, which is helped by alignment of the WSLs in each miniband,
and hence are strongly coupled leading to enhanced THz emission.
Increasing~$F$ partially leads to further localization, as in the
medium-field regime, but causes the WSLs to become misaligned,
thereby decreasing the Zener tunneling rate and causing a decrease
of $I$ with increasing~$F$. In this way, we can understand the
existence of these THz resonances, and, moreover, design SLs to
exploit these emission peaks; moreover our theory makes it evident
that, when second-order perturbation events become non-negligible,
wave functions become progressively delocalized thereby making the
SL an increasingly poor THz emitter.

In summary we have developed a clear model of electromagnetic
emission from a SL, based on single miniband dynamics in the low-
and medium-field regimes, and two-miniband dynamics in the high- and
ultrahigh-field regimes. Our results contain previous theories for
low- and medium-field dynamics as special cases and explain the
coincidence of enhanced Zener tunneling rates and
experimentally-observed THz emission peaks. We have excellent
quantitative agreement with experiment, subject to choosing two
fitting parameters by using least squares fitting method, and we
have a clear explanation of when and why THz emission becomes poor
in the ultrahigh applied electric field regime. In addition to the
value of our model as a design tool for SL THz emitters, our
perturbation theory approach is remarkably successful in explaining
electromagnetic radiation features in recent SL experiments due to
confinement or coupling between minibands within the conduction
band. Our approach also underscores the value of a perturbation
theory approach, as opposed to exact or numerical means, to
understand the underlying physics.

\acknowledgments We gratefully acknowledge financial support from
the National Natural Science Foundation of China (No.~60321003)
and National Basic Research Program of China, and BCS also
acknowledges support from iCORE, NSERC, and CIAR.

\end{document}